# Vote Elicitation: Complexity and Strategy-Proofness

**Vincent Conitzer** and **Tuomas Sandholm**
{conitzer, sandholm}@cs.cmu.edu

**Abstract**

Preference elicitation is a central problem in AI, and has received significant attention in single-agent settings. It is also a key problem in multiagent systems, but has received little attention here so far. In this setting, the agents may have different preferences that often must be aggregated using voting. This leads to interesting issues because what, if any, information should be elicited from an agent depends on what other agents have revealed about their preferences so far. In this paper we study effective elicitation, and its impediments, for the most common voting protocols. It turns out that in the Single Transferable Vote protocol, even knowing when to terminate elicitation is $\mathcal{N}P$-complete, while this is easy for all the other protocols under study. Even for these protocols, determining how to elicit effectively is $\mathcal{N}P$-complete, even with perfect suspicions about how the agents will vote. The exception is the Plurality protocol where such effective elicitation is easy. We also show that elicitation introduces additional opportunities for strategic manipulation by the voters. We demonstrate how to curtail the space of elicitation schemes so that no such additional strategic issues arise.

## 1 Introduction

Preference elicitation is a central problem in AI. To build a bot that acts intelligently on behalf of any type of agent (a human, a corporation, a software agent, etc.), the bot needs to know about the agent's preferences. However, the bot should only elicit pertinent preference information from the agent because determining and expressing preferences can be arduous. Significant work has been done on selective preference elicitation (e.g., (Boutilier *et al.* 1997; Vu & Haddawy 1998; Chajewska, Koller, & Parr 2000)).

Preference elicitation is also a key problem in multiagent systems, but has received little attention so far.[1] The agents may have different preferences over the set of *candidates* that the agents must collectively choose among (e.g., potential presidents, joint plans, resource allocations, task allocations, etc.). The most general method for aggregating preferences is voting.[2] In traditional voting, each voter is asked for its complete preferences. We observe that intelligently eliciting preferences from the voters can allow the voting protocol to determine the outcome well before all of the preferences have been elicited. This is desirable for any of several reasons: 1) it can be costly for an agent to determine its own preferences (e.g., computationally (Sandholm 1993;

[1] A notable exception is bid elicitation in combinatorial auctions (Conen & Sandholm 2001).

[2] Voting mechanisms have been used also for computational agents (e.g. (Ephrati & Rosenschein 1991; Ephrati & Rosenschein 1993)).

Larson & Sandholm 2001)), 2) communicating the preferences introduces overhead (network traffic, traveling to the voting booth to vote, traveling door to door to collect votes, etc.), and 3) less preference revelation is desirable due to privacy reasons.

Attempting to efficiently elicit preferences leads to interesting issues in the voting context because what, if any, information should be elicited from an agent depends on what other agents have revealed about their preferences so far. The goal here is to determine the right outcome while eliciting a minimal amount of preference information from the voters. The most effective elicitation schemes make use of *suspicions* about the agents' preferences. Such suspicions can be the result of votes in previous elections, an understanding of the candidates in the election, an understanding of how each agent relates to each candidate, etc. To see how suspicions may help the elicitation process, consider a simple election with two candidates. If the elicitor knew beforehand which agents would vote for the eventual winner, simply querying enough of those voters would suffice.

In this paper we analyze the possibility of effective vote elicitation, and demonstrate two categories of impediments. First, optimal elicitation can be computationally complex. In Section 4 we show that even determining whether enough information has been elicited is $\mathcal{N}P$-complete for some voting protocols. In Section 5 we show that for most of the voting protocols, determining an efficient elicitation policy is $\mathcal{N}P$-complete (even with perfect suspicions). Second, in Section 6 we show that in various ways, elicitation can introduce additional opportunities for strategically manipulating the election. We then show how to avoid such problems by curtailing the space of elicitation schemes.

## 2 Common voting protocols

In this section we summarize the common voting protocols that we analyze. We consider elections with $m$ candidates and $n$ voters (agents). A voting protocol defines a function from the set of all possible combinations of votes to the set of candidates, the *winner determination function*. We now review the most common protocols in use, all of which will be studied in this paper.

- *Plurality.* Each candidate receives one point for each voter that ranked it first. The candidate with the most points wins.

- *Borda.* For each voter, a candidate receives $m-1$ points if it is the voter's top choice, $m-2$ if it is the second choice, ..., 0 if it is the last. The candidate with the most points wins.

- *Copeland (aka. Tournament).* The protocol simulates a pairwise election for each pair of candidates in turn (in a pairwise election, a candidate wins if it is preferred over the other candidate by more than half of the voters). A candidate gets 1 point if it defeats an opponent, 0 points if it draws, and -1 points if it loses. The candidate with the most points wins.

- *Maximin.* A candidate's *score* in a pairwise election is the number of voters that prefer it over the opponent. A candidate's number of points is its lowest score in any pairwise election. The candidate with the most points wins.

- *Single Transferable Vote (STV).* The winner determination process proceeds in rounds. In each round, a candidate's score is the number of voters that rank it highest among the remaining candidates, and the candidate with the lowest score drops out. The last remaining candidate wins. (The name comes from the fact that a vote *transfers* from its top remaining candidate to the next highest remaining candidate when the former drops out.)

- *Approval.* Each voter labels each candidate as either approved or disapproved. The candidate that is approved by the largest number of voters wins.

## 3 Definition of elicitation

In this section, we formally define elicitation. We distinguish between *full elicitation*, where the entire vote is elicited from every agent; *coarse elicitation*, where upon querying an agent the elicitor always asks for the agent's entire vote; and *fine elicitation* where this need not be the case (for example, an agent may be asked only what its most preferred candidate is). We formalize elicitation policies as trees.

**Definition 1** *A* coarse elicitation tree *is a tree with the following properties:*

- *Each nonleaf node $v$ is labeled with an agent $a_v$ to be queried.*
- *Each nonleaf node $v$ has a child for each of the possible votes of agent $a_v$.*
- *On each path from the root to a leaf, each agent occurs at most once.*

This tree determines how the elicitation will proceed for any combination of votes by the agents. The elicitor starts at the root. At each node, it queries the corresponding agent, and subsequently moves to the child corresponding to the obtained vote. We say the tree is *valid* for a protocol when for each leaf, given the votes corresponding to the path from the root to that leaf, the election's outcome is determined.

**Definition 2** *A* fine elicitation tree *is a tree with the following properties:*

- *Each nonleaf node $v$ is labeled with an agent $a_v$ to be queried, a subset of that agent's possible votes $S_v$ (the ones consistent with $a_v$'s responses so far), and a query to be asked at that node. The query is given by a partition $\mathcal{T}_v$ of $S_v$; once the query is answered, one element of $\mathcal{T}_v$ is the set of remaining consistent votes.*
- *Each nonleaf node $v$ has a child for each element of $\mathcal{T}_v$.*

- *Given a nonleaf $v$, if $a_v$ does not occur anywhere else on the path from the root to $v$, then $S_v$ is the set of all possible votes by $a_v$; otherwise, consider the node $w$ closest to $v$ on that path with $a_v = a_w$. The element of $\mathcal{T}_w$ corresponding to $w$'s child on the way to $v$ must equal $S_v$.*
- *Each partition $\mathcal{T}_x$ has at least 2 elements.*

The interpretation is as follows. Each node still corresponds to a query to the corresponding agent. A subset at a node is the set of the agent's possible votes that are consistent with its responses so far. The partition indicates the various ways in which this set may be reduced through the query.

We say the tree is *valid* for a protocol if for each leaf, the outcome of the election is determined by the responses to the queries on the path to that leaf. From now on, we only consider valid trees.

Our model of elicitation is very general. It can be used to represent intuitively reasonable queries as well as baroque ones such as *"Is it true that $a$ is your most preferred candidate or that you prefer $b$ to $c$?"* (which could impose a computational burden on the voter disproportionate to the fact that it is only one query). Reasonable fine elicitation policies will have some restriction on the types of $\mathcal{T}_v$ allowed. Also, elicitation trees can be extremely large. Therefore, it can be unreasonable to expect the elicitor to use this explicit representation for its elicitation policy, much less to do an exhaustive search over these trees to find one that minimizes the number of queries (for example, in the average case). Nevertheless, each well-defined elicitation policy corresponds to an elicitation tree, and hence elicitation trees are useful tools for analysis.

## 4 Hardness of terminating elicitation

Any sensible elicitation policy would need to be able to determine when it can safely terminate. Otherwise, there would be no benefit from elicitation. In this section, we first show that for the STV protocol, it can be hard to determine when the elicitation process can terminate. Then we show that this is easy for the other voting protocols.

**Definition 3 (ELICITATION-NOT-DONE)** *We are given a set of votes $S$, a number $t$ of votes that are still unknown, and a candidate $h$. We are asked whether there is a way to cast the $t$ votes so that $h$ will not win.*

In order to prove our hardness result, we make use of the following result from the literature on the difficulty of manipulating an election.

**Definition 4 (EFFECTIVE-PREFERENCE)** *We are given a set of votes $S$ and a candidate $c$. One vote is not yet known. Is there a way to cast the last vote that makes $c$ win?*

**Theorem 1 (Known)** *For the STV protocol, EFFECTIVE-PREFERENCE is $\mathcal{NP}$-complete, even under the restriction that at least one of the votes in $S$ puts $c$ in the top spot.*

(The restriction is of little interest in itself, but we will use it for our reduction.)

**Proof**: This was proven in (Bartholdi & Orlin 1991). ∎

**Theorem 2** *For the STV protocol, ELICITATION-NOT-DONE is $\mathcal{NP}$-complete, even when $t = 1$.*[3]

**Proof**: We reduce an arbitrary instance of EFFECTIVE-PREFERENCE (with the restriction that at least one of the votes in $S$ puts $c$ at the top) to an instance of ELICITATION-NOT-DONE as follows. In the EFFECTIVE-PREFERENCE instance, let the candidate set be $C_{EP}$ and the set of given votes $S_{EP}$. Then, in our ELICITATION-NOT-DONE instance, the candidate set is $C_{EP} \cup \{h\}$. The known (elicited) set of votes $S$ includes all the votes from $S_{EP}$, where $h$ is appended to these votes at the bottom – with the exception that one of the votes with $c$ at the top inserts $h$ into the second place (right behind $c$). Additionally, $S$ includes $|S_{EP}|$ additional votes which place $h$ in the top spot and rank the other candidates in whichever order. Finally, we set $t = 1$. We prove the instances are equivalent by making the following observations. First, $h$ will always survive until the last round as it has almost half the votes at the start. Second, if there exists a way for the last vote to be cast such that $h$ does not win the election, we may assume that this vote places $h$ at the bottom, since if this vote ever transferred to $h$, $h$ would win the election as it would hold more than half the votes. Third, $h$ will not win the election if and only if it faces $c$ in the last round (if $c$ gets eliminated, the vote that ranks $h$ right below $c$ would transfer to $h$ and $h$ would win the election; on the other hand, $c$ is ranked above $h$ in all the votes that do not put $h$ at the top, so $c$ would win the last round). Fourth, as long as $c$ remains in the election, the score of each candidate (besides $h$) in each round before the last will be exactly the same as the corresponding score in the EFFECTIVE-PREFERENCE instance (if we give the same value to the unknown vote in both instances). This follows from the fact that in this case, no vote will ever transfer to or from $h$ and the relevant votes are identical otherwise. It follows that the remaining vote can be cast in such a way as to lead $c$ to the final round if and only if the remaining vote in the EFFECTIVE-PREFERENCE instance can make $c$ win the election. But then, by our third observation, the instances are equivalent. ∎

Theorem 3 applies to both fine and coarse elicitation because in both the elicitor might end up in a situation where it has elicited some votes completely and others not at all.

For all the other voting protocols discussed in this paper, it can be determined in polynomial time whether elicitation can be terminated. This is done by running a simple greedy algorithm (omitted due to limited space) that is guaranteed to find $t$ votes that make $h$ not win if such $t$ votes exist.

## 5  Hardness of deciding which votes to elicit

The elicitor could use its suspicions about how the agents will vote to try to design the elicitation policy so that few queries are needed. The suspicions could be represented by a joint prior distribution over the agents' votes. It is not too surprising that in this general setting, computational complexity issues arise with regard to optimal elicitation, be-

---

[3]In all $\mathcal{NP}$-completeness proofs, we only prove $\mathcal{NP}$-hardness because proving that the problem is in $\mathcal{NP}$ is trivial.

---

cause the number of probabilities in a general joint prior distribution is $(m!)^n$. Given that this is an impractically large amount of information to generate (and to input into an elicitor bot), it is reasonable to presume that the language the elicitor uses to express its suspicions is not fully expressive. With such a restricted language, one might hope that the optimal elicitation problem is tractable. However, this turns out not to be the case! We show that if this language even accomodates as little as degenerate distributions (all the probability mass on a single vote), determining an optimal coarse elicitation policy is hard. In other words, it is hard even with perfect suspicions. We define the effective elicitation problem with perfect suspicions as follows:

**Definition 5 (EFFECTIVE-ELICITATION)** *We are given a set of votes $S$ and a number $k$. We are asked whether there is a subset of $S$ of size $\leq k$ that decides the election constituted by the votes in $S$.*

All the reductions in this section will be from 3-COVER.

**Definition 6 (3-COVER)** *We are given a set $U$ of size $3q$ and a collection of subsets $\{S_i\}_{1 \leq i \leq r}$ of $U$ (where $r > q$), each of size 3. We are asked if there is a cover of $U$ consisting of $q$ of the subsets.*

**Theorem 3** *For the Approval protocol, EFFECTIVE-ELICITATION is NP-complete.*

**Proof**: We reduce an arbitrary 3-COVER instance to the following EFFECTIVE-ELICITATION instance. The candidate set is $U \cup \{w\}$. The votes are as follows. For every $S_i$ there is a vote approving $S_i \cup \{w\}$. Additionally, we have $r - 2q + 2$ votes approving only $\{w\}$, for a total of $2r - 2q + 2$ votes. Finally, we set $k = r - q + 2$. We claim the problem instances are equivalent. First suppose there is a 3-cover. Then we elicit all the votes that approve only $w$, and the votes that correspond to sets in the cover, for a total of $k$ votes. Then $w$ is $r - q + 1$ points ahead of all other candidates, with only $r - q$ votes remaining. Hence there is an effective elicitation. On the other hand, suppose there is no 3-cover. Then eliciting $k$ votes will always give one of the candidates in $U$ at least 2 votes, so that $w$ can be at most $r - q$ points ahead of this candidate. Hence, with $r - q$ votes remaining, the election cannot possibly be decided. So there is no effective elicitation. ∎

**Theorem 4** *For the Borda protocol, EFFECTIVE-ELICITATION is NP-complete.*

**Proof**: We reduce an arbitrary 3-COVER instance to the following EFFECTIVE-ELICITATION instance. The candidate set is $U \cup \{w\} \cup B$ where $B = \{b_1, b_2, \ldots, b_{64r^2}\}$. The votes are as follows. For each $S_i$ there is a vote ranking the candidates $(B/2, U - S_i, B/2, S_i, w)$, where the occurrence of a set in the ranking signifies all of its elements in whichever order, and $B/2$ signifies some subset of $B$ containing half its elements. Finally, there are $4r - 2q - 2$ votes that rank the candidates $(w, b_1, \ldots, b_{8r^2}, u_1, \ldots, u_{3q}, b_{8r^2+1}, \ldots, b_{64r^2})$, and another $4r - 2q - 2$ that rank them

$(w, b_{64r^2}, \ldots, b_{56r^2+1}, u_{3q}, \ldots, u_1, b_{56r^2}, \ldots, b_1)$. Let $g = 8r - 4q - 4$, so that we have a total of $g + r$ votes. Also, let $l = 64r^2 + 3q$, which is the number of points a candidate gets for being in first place. Finally, we set $k = g + q$. We claim the problem instances are equivalent. First suppose there is a 3-cover. We elicit all the votes that put $w$ on top, and the votes that correspond to sets in the cover, for a total of $k$ votes. Even after eliciting just the ones that put $w$ on top, $w$ is more than $\frac{gl}{2} \geq 2rl$ (since $g \geq 4r$) points ahead of all the elements of $B$, and with only $r$ votes remaining it is impossible to catch up with $w$ for anyone in $B$. For a given element $u$ of $U$, the votes that put $w$ on top result in a net difference of $g(8r^2 + \frac{3}{2}q + \frac{1}{2})$ points between $w$ and $u$. Of the $q$ remaining elicited votes, precisely $q - 1$ placed $u$ ahead of half the elements of $B$, so the net difference in points between $w$ and $u$ most favorable to $u$ arising from these would be $-(q-1)(32r^2 + 3q)$. Finally, the vote that put $u$ below all the elements of $b$ might contribute another $-3$. Adding up all these net differences, we find that $w$ is ahead by at least $64r^3 - 64qr^2 + 12qr - 9q^2 + 4r - 5q - 5$ points. On the other hand, the maximum number of points $u$ could gain on $w$ with the remaining number of votes is $(r - q)(64r^2 + 3q) = 64r^3 - 64qr^2 + 3qr - 3q^2$. It is easily seen that the second expression is always smaller, and hence $w$ is guaranteed to win the election. So there is an effective elicitation. On the other hand, suppose there is no 3-cover. First, we observe that $w$ will always win the election - we have already shown that the votes that put $w$ on top guarantee it does better than any element of $B$. For any element $u$ of $U$, even if $u$ is always placed above all the other votes in $U$ in the $r$ votes corresponding to the $S_i$, it will still only gain $r(32r^2 + 3q)$ points on $w$ here, which is fewer than the $g(8r^2 + \frac{3}{2}q + \frac{1}{2})$ votes it loses on $w$ with the other votes (since $g \geq 4r$). So we can only hope to guarantee that $w$ wins. Now, if there is an elicitation that guarantees this, there is also one that elicits all the $g$ votes that put $w$ on top, since replacing one of the other votes with such a vote in the elicitation never hurts $w$'s relative performance to another candidate. But in such an elicitation, there is at least one candidate $u$ in $U$ that is never ranked below all the elements of $B$ in the $q$ votes elicited that put $w$ at the bottom, since there is no 3-cover. Let us investigate how many points $w$ may be ahead of $u$ after eliciting these votes. Again, the votes that put $w$ on top result in a net difference of $g(8r^2 + \frac{3}{2}q + \frac{1}{2})$ points. In the scenario most favorable to $w$, $u$ would only gain $q(32r^2 + 4)$ points with the other $q$ votes. Adding this up, $w$ is ahead by at most $64r^3 - 64qr^2 - 32r^2 + 12qr - 6q^2 + 4r - 6q - 2$ after the elicitation. The maximum number of points $u$ could gain on $w$ with the remaining number of votes is still $64r^3 - 64qr^2 + 3qr - 3q^2$. It is easily seen that the second expression is always larger, so we cannot guarantee that $w$ wins. So there is no effective elicitation. ∎

**Theorem 5** *For the Copeland protocol,* EFFECTIVE-ELICITATION *is NP-complete.*

**Theorem 6** *For the Maximin protocol,* EFFECTIVE-ELICITATION *is NP-complete.*

The proofs are omitted due to limited space – the ideas are similar to those used for the Approval and Borda protocols.

So far we have shown that determining an effective elicitation policy is hard for most of the voting protocols, and that for the STV protocol even knowing when to terminate is hard. The remaining protocol is Plurality, where it is easy to elicit effectively given perfect suspicions (start eliciting the winner's votes first; if all of them have been elicited and termination is still not possible, elicit votes in a round-robin manner, one for each non-winning candidate (as long as it has votes left), until the elicitation can terminate).

## 6 Strategy-proofness of elicitation

We now turn to strategic issues that may be introduced into a voting protocol by an elicitation process. Elicitation may reveal information about other agents' votes to an agent, which the agent may use to change its vote strategically. This is undesirable for two reasons. First, it gives agents that are elicited later an unfair advantage, causing the protocol to put undue weight on their preferences. Second, it leads to less truthful voting by the agents. This is undesirable because, while voting protocols are designed to select a socially desirable candidate if agents vote truthfully, untruthful voting can lead to a reduction in the social desirability of the outcome. We demonstrate how such strategic issues may arise, and then suggest avenues to circumvent them. However, these avenues entail restricting the space of possible elicitations, causing a reduction in the potential savings from elicitation.

To analyze strategic interactions, we need some tools from game theory. To bring the voting setting into the framework of noncooperative game theory, we assume that agent $i$'s preferences are defined by its *type* $\theta_i$; the agent gets utility $u_i(\theta_i, c)$ if candidate $c$ wins. We first define a game:

**Definition 7** *In a* (normal form) game*, we are given a set of agents $A$; a set of types $\Theta_i$ for each agent $i$; a commonly known prior distribution $\phi$ over $\Theta_1 \times \Theta_2 \times \ldots \times \Theta_{|A|}$; a set of strategies $\Sigma_i$ for each agent $i \in A$; a set of outcomes $O$ (candidates in the case of voting); an outcome function $o : \Sigma_1 \times \Sigma_2 \times \ldots \times \Sigma_{|A|} \rightarrow O$; and a utility function $u_i : \Theta_i \times O \rightarrow \Re$ for each agent $i \in A$.*

An agent knows its own type and can thus let its strategy depend on its type according to a function $f_i : \Theta_i \rightarrow \Sigma_i$. We also need a notion of how an agent would play a game strategically. This may depend on how others play.

**Definition 8** *A strategy function profile $(f_1, f_2, \ldots, f_{|A|})$ is a Bayes-Nash equilibrium (BNE), if for each agent $i \in A$, each $\theta_i \in \Theta_i$, and each strategy $\sigma_i \in \Sigma_i$, $E_\phi(u_i(\theta_i, o(f_1(\theta_1), f_2(\theta_2), \ldots, f_i(\theta_i), \ldots, f_{|A|}(\theta_{|A|})))|\theta_i) \geq E_\phi(u_i(\theta_i, o(f_1(\theta_1), f_2(\theta_2), \ldots, \sigma_i, \ldots, f_{|A|}(\theta_{|A|})))|\theta_i)$ (that is, each $f_i$ chooses, for each $\theta_i$, a strategy that maximizes $i$'s expected utility given the other players' $f_j$s).*

We are now ready to state our results.

### 6.1 Coarse elicitation

First we show that coarse elicitation may lead to strategic manipulations when it reveals even slightly more than just the fact that the agent is being elicited.

**Theorem 7** *Consider a coarse elicitation protocol that has the property that when an agent's type is elicited, the agent knows how many agents have had their types elicited so far. Then the following properties can hold simultaneously:*

- *truthful voting is a BNE in the corresponding full elicitation case,*
- *truthful voting is not a BNE here,*
- *the elicitation policy is optimized to finish as quickly as possible on average given the distribution over the agents' types (presuming the agents vote truthfully), and*
- *in a given BNE, an agent may vote differently depending on what the other agents' types end up being.*

**Proof**: Consider an Approval election with 3 voters, $i$, $j$ and $k$, and 3 candidates, $a$, $b$, and $c$. Ties are broken randomly. Define truth-telling to mean approving all candidates that give you utility $\geq \frac{1}{2}$. Ties are broken randomly. Agents' types are independent and the distributions are as follows. With probability $\frac{1}{2}$, $i$ has utility 1 for $c$, and utility 0 for $a$ and $b$; with probability $\frac{1}{2}$, it has utility 1 for $a$, and utility 0 for $b$ and $c$. With probability $\frac{1}{2}$, $j$ has utility 1 for $c$, and utility 0 for $a$ and $b$; with probability $\frac{1}{2}$, it has utility 1 for $b$ and $c$, and utility 0 for $a$. It is easy to see that truth-telling is always an optimal strategy for $i$ and $j$. With probability 1, $k$ has utility 1 for $a$, $\frac{1}{4}$ for $b$, and 0 for $c$. For $k$ not to approve $c$, and to approve $a$, is always optimal. In the full elicitation case, should $k$ approve $b$? If $j$ has its first type, it makes no difference. What if $j$ has its second type? If $i$ has its first type, approving $b$ leads to a tie between $b$ and $c$, and (expected) utility $\frac{1}{8}$; not approving $b$ leads to a victory for $c$ and utility 0. If $i$ has its second type, approving $b$ leads to a tie between $a$ and $b$ and utility $\frac{5}{8}$; not approving $b$ leads to a victory for $a$ and utility 1. Hence, in the full elicitation case, given that we are in a case where it matters whether $k$ approves $b$, approving $b$ gives utility $\frac{3}{8}$, and not approving $b$ gives utility $\frac{1}{2}$; so not approving $b$ is optimal. Thus, truth-telling is a BNE here.

For the coarse elicitation case, we first design a policy that is optimal with respect to the agents' type distributions. Query $Q(l)$ asks voter $l$ which candidates it approves. Then an optimal elicitation protocol is

*1*: first ask $Q(i)$;
*2a*: if the answer was $\{c\}$, ask $Q(j)$;
*2b*: otherwise, ask $Q(k)$;
*3*: if do not know the winner yet, query the last voter.

To show optimality with respect to the type distribution, assume the agents reply truthfully. If $i$ has its first type, and $j$ its first, we finish after *2a*, in 2 steps. If $i$ has its second type, we finish after *2b*, in 2 steps. But these are the only cases in which we can hope to finish in only two steps, so the protocol is optimal.

Now, if $k$ is queried second, this implies to it that $i$ is of its second type, and it is motivated to answer truthfully. But if $i$ is queried third, this implies to it that $i$ is of its first type, and $j$ of its second type; and $k$ is motivated to lie and approve $b$. So truth-telling is not a BNE here. ∎

However, if the elicitation reveals no information to the agent being elicited (beyond the fact that the agent is being elicited), then elicitation does not introduce strategic issues:

**Theorem 8** *Consider a coarse elicitation protocol which manages to reveal nothing more to the agent than whether or not his type is elicited. Then, the set of BNEs is the same as in the corresponding full elicitation voting game.*[4]

**Proof**: We claim that the normal form of the game is identical to that in the full elicitation setting; this implies the theorem. Obviously, the $\Theta_i$, the $u_i$, and $\phi$ remain the same. Now consider the $\Sigma_i$. Because no information is revealed upon elicitation, the voter cannot condition its response on anything but its type, as in the full elicitation case. That is, each agent need only decide on the one vote that it will always cast if it is elicited. Hence, the strategy set of an agent is simply the space of votes, as it is in the full elicitation case. Finally, by our requirement that this elicitation produces the same outcome as full elicitation, $o$ is the same. ∎

It is an interesting open problem how to design an elicitation protocol that reveals no information about how many agents have had their types elicited so far. This seems difficult because any protocol will at least betray the real time at which an agent is queried.

### 6.2 Fine elicitation

We now show that fine elicitation can lead to additional strategic issues even if no unnecessary information is revealed to the agents.

**Theorem 9** *In a fine elicitation protocol, the following properties can hold simultaneously:*

- *the protocol reveals no information to any agent except the queries to the agent and the order of those queries,*
- *truthful voting is a BNE in the corresponding full elicitation case,*
- *truthful voting is not a BNE here,*
- *the elicitation policy is optimized to finish as quickly as possible on average given the distribution over the agents' types (presuming the agents vote truthfully), and*
- *in a given BNE, an agent may vote differently depending on what the other agents' types end up being.*

**Proof**: Consider an Approval election with 2 voters, $i$ and $j$, and 3 candidates, $a$, $b$, and $c$. Define truth-telling to mean approving all candidates that give you utility $\geq \frac{1}{2}$. Ties are broken randomly. Agents' types are independent and the distributions are as follows. With probability $\frac{1}{2}$, $i$ has utility 1 for $b$ and $c$, and utility 0 for $a$; with probability $\frac{1}{2}$, it has utility 1 for $a$ and $b$, and utility 0 for $c$. It is easy to see that truth-telling is always an optimal strategy for $i$. With probability 1, $j$ has utility 1 for $a$, $\frac{3}{4}$ for $b$, and 0 for $c$. For $j$ not to approve $c$, and to approve $a$, is always optimal. In the full elicitation case, should $j$ approve $b$? If $i$ has its first type, approving $b$ leads to victory for $b$ and a utility of $\frac{3}{4}$; not approving $b$ leads to a 3-way tie and utility of $\frac{7}{12}$. If $i$ has its second type, approving $b$ leads to a 2-way tie between $a$ and $b$ and utility $\frac{7}{8}$; not approving $b$ leads to a victory for $a$ and utility 1. Hence, in the full elicitation case, approving $b$ gives utility $\frac{13}{16}$, and not approving $b$ gives utility $\frac{19}{24}$; so approving $b$ is optimal. Thus, truth-telling is a BNE here.

---

[4]For the game-theoretically inclined, we observe that some of the BNEs in the coarse elicitation case are not subgame perfect. These equilibria are unstable in the full elicitation case as well.

For the fine elicitation case, we first design a policy that is optimal with respect to the agents' type distributions. The natural restriction here is to allow only the following type of query: query $Q(k, d)$ asks voter $k$ if it approves candidate $d$. Then an optimal elicitation protocol is

*1*: first ask $Q(i, a)$;
*2a*: if the answer was 'no', ask $Q(i, b)$; $Q(j, b)$; $Q(j, c)$;
*2b*: otherwise, ask $Q(j, a)$; $Q(i, b)$; $Q(j, b)$; $Q(i, c)$.

*3*: if we do not know the winner yet, ask the remaining queries.

To show optimality with respect to the type distribution, assume the agents reply truthfully. If $i$ has its first type, we finish after *2a*, in 4 steps; if $i$ has its second type, we finish after *2b*, in 5 steps. This is optimal.

Now, if the first query to $j$ is $Q(j, b)$, this implies to it that $i$ is of its first type and it is motivated to answer truthfully. But if the first query to $j$ is $Q(j, a)$, this implies to it that $i$ is of its second type; and $i$ is motivated to lie and not approve $b$. So truth-telling is not a BNE here. ∎

Finally, we show that with a certain restriction on elicitation policies, we can guarantee that fine elicitation does not introduce any strategic effects.

**Definition 9** *A fine elicitation policy is* nondivulging *if the next query to an agent (if it comes) depends only on that agent's own responses to previous queries. (Whether or not the next query is asked can depend on the agent's* and the other agents' *responses to queries so far.)*

**Theorem 10** *Consider a fine elicitation protocol which manages to reveal nothing more to the agent than the queries to the agent and the order of those queries. If the elicitation policy is nondivulging, then the set of BNEs is the same as in the full elicitation voting game.*

**Proof**: We claim that the normal form of the game is identical to that in the full elicitation setting; this implies the theorem. Obviously, the $\Theta_i$, the $u_i$, and $\phi$ remain the same. Now consider the $\Sigma_i$. Because the agent knows the first query to it (if it comes), it can determine its response up front. The next query (if it comes) can only depend on this response, so the agent knows it, and can prepare a response to it up front as well; and so on. So, in this setting, we can define the agent's strategy to be this entire sequence of responses. But this sequence correponds to exactly one vote in the full elicitation case.[5] Hence, the strategy set of an agent is simply the space of votes, as it is in the full elicitation case. Finally, by our requirement that this elicitation produces the same outcome as full elicitation, $o$ is the same. ∎

While a restriction to nondivulging elicitation policies avoids introducing additional strategic effects, it can reduce the efficiency of elicitation.

## 7 Conclusion and future research

Preference elicitation is a central problem in AI, and has received significant attention in single-agent settings. Preference elicitation is also a key problem in multiagent systems, but has received little attention here. In this setting, the agents may have different preferences that often must be aggregated using voting. This leads to interesting issues because what, if any, information should be elicited from an agent depends on what other agents have revealed about their preferences so far.

In this paper we studied effective elicitation for the most common voting protocols. It turned out that for the STV protocol, even knowing when to terminate elicitation is $\mathcal{NP}$-complete, while this is easy for all the other protocols. Even for these protocols, determining how to elicit effectively is $\mathcal{NP}$-complete, even with perfect suspicions about how the agents will vote. The exception is the Plurality protocol where such effective elicitation is easy.

Our results on strategy-proofness showed that in general settings, elicitation introduces additional opportunities for strategic manipulation of the election by the voters. We demonstrated how to curtail the space of elicitation schemes so that no such additional strategic issues arise.

Future research includes studying elicitation policies that choose the right outcome with high *probability* rather than with certainty. It also includes designing new voting protocols that combine the computational ease of elicitation in the Plurality protocol with the expressiveness of the other protocols. Finally, it would be interesting to study specific fine elicitation schemes in more detail.

---

[5]By our definition of a fine elicitation policy, no queries are asked that would enable an agent to express inconsistent (e.g., cyclical) preferences.